\documentclass[%
 aip,
 amsmath,amssymb,
 reprint,
]{revtex4-1}
\setcitestyle{super}
\usepackage{graphicx}
\usepackage{dcolumn}
\usepackage{bm}
\usepackage{epstopdf}

\usepackage[utf8]{inputenc}
\usepackage[T1]{fontenc}
\usepackage{mathptmx}

\newcommand*{\citen}{}% generate error, if `\citen` is already in use
\DeclareRobustCommand*{\citen}[1]{%
  \begingroup
    \romannumeral-`\x % remove space at the beginning of \setcitestyle
    \setcitestyle{numbers}%
    \cite{#1}%
  \endgroup
}

\begin{document}

\preprint{AIP/123-QED}

\title[Hybrid Magnetoacoustic Metamaterials for Ultrasound Control]{Hybrid Magnetoacoustic Metamaterials for Ultrasound Control}

\author{O. S. Latcham}
    \affiliation{University of Exeter, Stocker Road, Exeter, EX4 4QL, United Kingdom}
\author{Y. I. Gusieva}
    \affiliation{Igor Sikorsky Kyiv Polytechnic Institute, 37 Prosp. Peremohy, Kyiv, 03056, Ukraine}
\author{A. V. Shytov}
    \affiliation{University of Exeter, Stocker Road, Exeter, EX4 4QL, United Kingdom}
\author{O. Y. Gorobets}
    \affiliation{Igor Sikorsky Kyiv Polytechnic Institute, 37 Prosp. Peremohy, Kyiv, 03056, Ukraine}
\author{V. V. Kruglyak}
    \email{V.V.Kruglyak@exeter.ac.uk}
    \affiliation{University of Exeter, Stocker Road, Exeter, EX4 4QL, United Kingdom}
\date{\today}

\begin{abstract}
%Propagation of acoustic waves may be magnetically 
%controlled via magnetoelastic coupling. 
%
We propose a class of metamaterials in which propagation of acoustic waves is controlled magnetically through magnetoelastic coupling. 
The metamaterials are formed by a periodic array of thin magnetic layers ('resonators') embedded in a non-magnetic matrix. 
Acoustic waves carrying energy through the structure hybridize with the magnetic modes of the resonators ('Fano resonance').
This leads to a rich set of effects, enhanced by Bragg scattering and being most pronounced when the magnetic resonance frequency is close to or lies within acoustic band gaps.
The acoustic reflection from the structure exhibits magnetically induced transparency and Borrmann effect. 
Our analysis shows that the combined effect of the Bragg scattering and Fano resonance may overcome the magnetic damping ubiquitous in realistic systems.
This paves a route towards application of such structures in wave computing and signal processing. 

%Magneto-acoustic resonators form promising nanostructures that may provide low-loss, non-volatile and reconfigurable magnetic computing elements. To analyse the feasibility of magneto-acoustic devices, we investigate a one-dimensional magneto-acoustic metamaterial formed by both a finite and semi-infinite number of magnetostrictive slabs sandwiched between a non-magnetic matrix. We demonstrate in the response function and resonant scattering that the magneto-elastic singularity hybridises with the background phononic dispersion. This combination optimises the magneto-elastic features at physical parameters when the Kittel modes are tuned to phononic band gaps. These structures show induced transmission and a Borrmann effect inside band gaps. We envision these characteristics will provide the strong tunability, as shown by a modulation coefficient, and high energy efficiency for use in future nanoscale magnonic devices.
\end{abstract}

\pacs{}% insert suggested PACS numbers in braces on next line

\maketitle 

Minimising energy loses in modern computing devices calls for unorthodox approaches to signal processing.\cite{Perrucci_2011, Liang_2019}
%As the number of computing devices rises, the deficit between their energy demand and the output of the energy production industry can be eased by tackling the inefficient heat loss in these devices.
For instance, proposals to employ spin waves\cite{Gurevich_1996} as a data carrier have lead to energy savings in non-volatile memory devices, promoting growth in the research area of magnonics.\cite{Kruglyak_2010}
%spin waves\cite{Gurevich_1996} have been proposed as a solution, whereby their use as data carriers could
However, these hopes are hampered by the short propagation distance of spin waves, caused by the magnetic damping.\cite{Krivoruchko_2015, Azzawi_2017} Magnetostrictive materials offer a route to circumvent this. Indeed, acoustic waves have longer attenuation lengths as compared to spin waves at the same frequencies. 
In magnetostrictive materials, acoustic waves can still couple to spin waves, forming hybrid magnetoacoustic waves.\cite{Kittel_1958, Bommel_1959, Dreher_2012, Callen_1965} Thus one regains the option of magnetic control and programmability, catering to the design of systems that evoke benefits of both acoustics and magnonics in terms of the energy efficiency.
\begin{figure}[ht!]
    \centering
    \includegraphics[width=85mm]{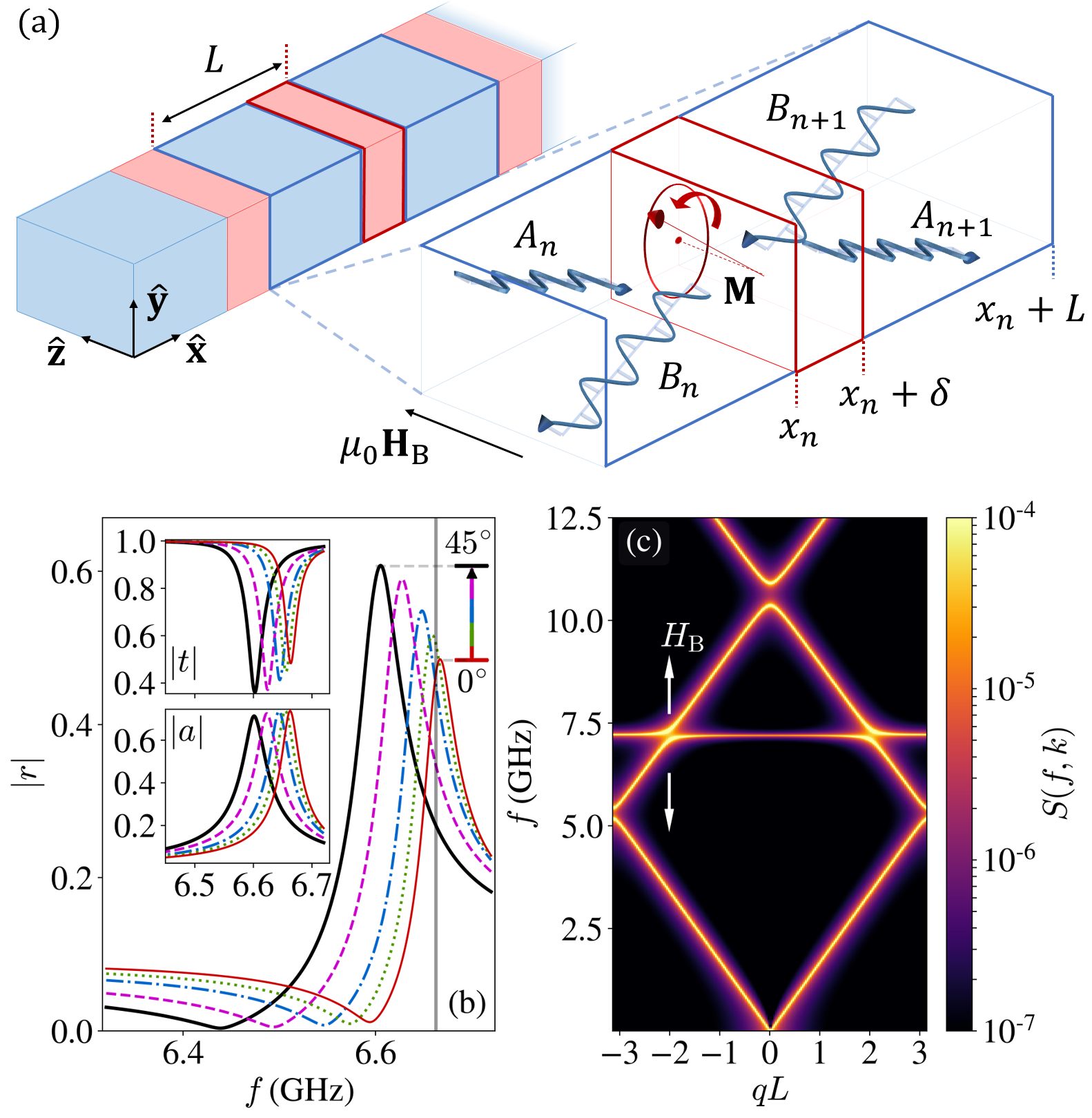}
    \caption{\label{fig:fig1}(a) The problem geometry is schematically shown. The metamaterial is formed by a 1D array of thin-film magnetoacoustic resonators embedded in a nonmagnetic matrix. Individual resonators scatter acoustic waves incident from both sides.  A bias magnetic field~$\mu_{0}{\bf H}_{\mathrm{B}}$ is applied in the resonator's plane.
    (b) The frequency dependence of the reflection coefficient, $r$, for incidence angles ranging from $0^{\circ}$ to $45^{\circ}$ is shown for an isolated Ni resonator in a silicon nitride matrix. The vertical line indicates the Kittel frequency for a field strength of $\mu_{0}H_{\mathrm{B}}=120$mT. The inset shows corresponding transmission, $t$, and absorption, $a$, coefficients. (c) The spectral function, $S(f, k)$, of acoustic waves in the metamaterial is shown. The frequency of the anticrossing is controlled by the bias magnetic field which is shown for a value of $\mu_{0}H_{\mathrm{B}} = 135$mT. 
    }
\end{figure} 

The recently studied magnetoacoustic devices\cite{Kamra_2015} and metamaterials\cite{Graczyk_2017} were typically formed using alternating magnetostrictive materials, so that the full acoustic and magnonic spectra were hybridized. To reduce the influence of the magnetic damping, we explored systems in which the magnetic loss was restricted to an isolated, thin-film magnetostrictive inclusion, hosting a single spin-wave mode, that of the ferromagnetic resonance (FMR).\cite{Latcham_2019} The FMR mode hybridized with acoustic waves only near the Kittel frequency,\cite{Gurevich_1996} which led to their resonant scattering in a magnetoacoustic version of the Fano resonance.\cite{Limonov_2017} The FMR mode's frequency and linewidth (and therefore the strength of the Fano resonance) were determined by the bias magnetic field and by the magnetic damping, respectively. %For a sizable quantitative experimental measurement we noted the system would have be carefully designed, introducing a figure of merit relating the decay channels associated with the magnetoacoustic mode,
Our analysis highlighted the need to enhance the (generally, weak) magnetoelastic interaction and to suppress the (generally, strong) magnetic damping, which was partly achieved by adopting an oblique incidence geometry. A question arises as to whether the effects of the magnetoelastic coupling could be enhanced even further due to Bragg scattering in magnetoacoustic metamaterials\cite{Graczyk_2017} formed by periodic arrays of magnetostrictive inclusions from Ref.~\citen{Latcham_2019}. 

In this Letter, we demonstrate that, by combining individual magnetoacoustic resonators into one-dimensional (1D) arrays, one can significantly enhance their effect on incident acoustic waves. The acoustic reflectivity of the structure exhibits a peak due to the magnetoacoustic Fano resonance. This peak's height and shape can be tuned at frequencies in the proximity of phononic band gaps. In particular, its behaviour near the two edges of a band gap exhibits a strong asymmetry, which is linked to the Borrmann effect.\cite{Novikov_2017} Inside the band gaps we identify behavior reminiscent of the magnetically induced transparency.\cite{Limonov_2017} These features of our prototypical structure could be employed to process acoustic signals and to readout magnetic information. 
\begin{figure}[t!]
    \centering
    \includegraphics[width=85mm]{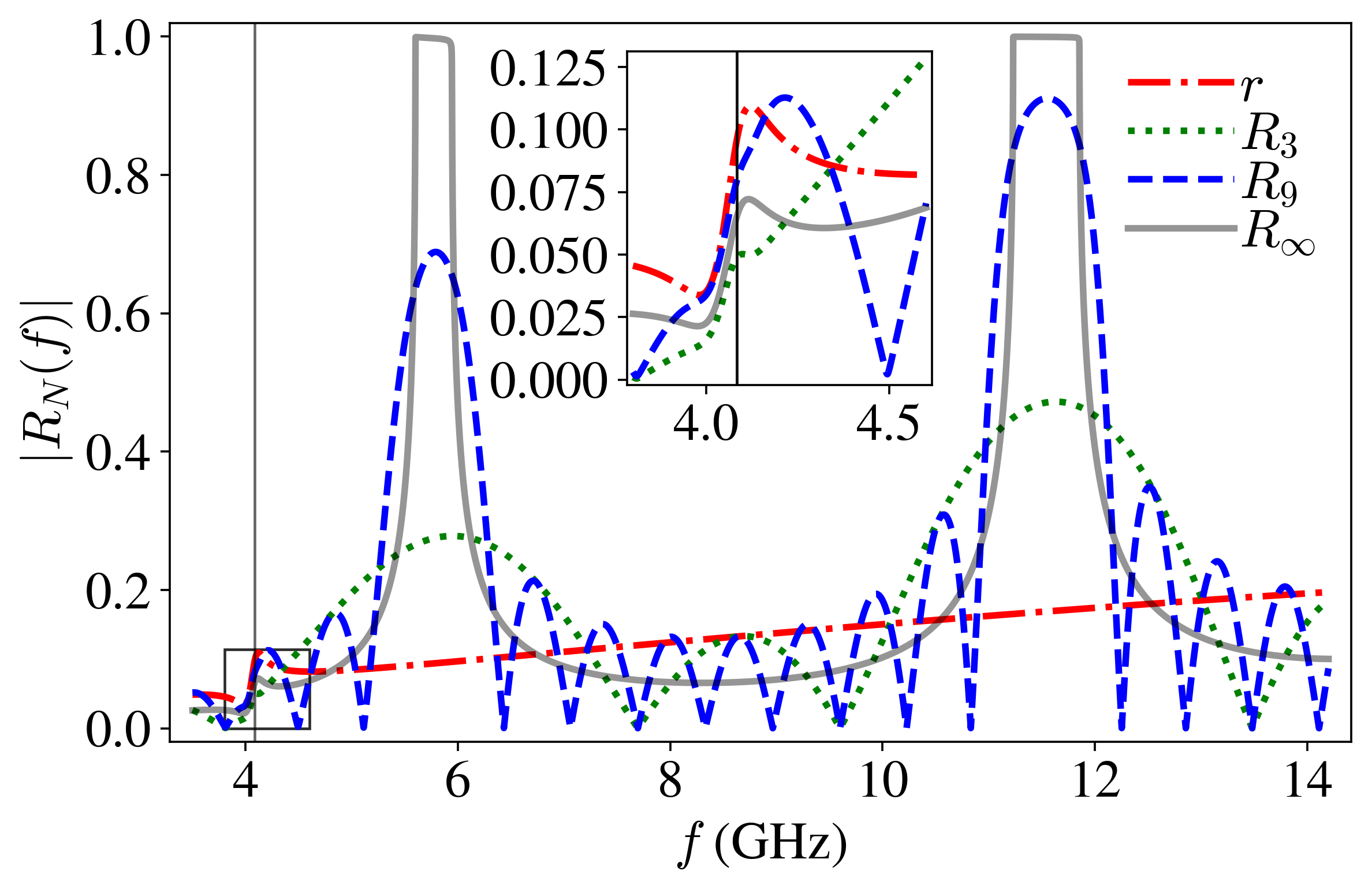}
    \caption{\label{fig:RnFigure} The frequency dependence of the reflection coefficient, $R_{N}$, calculated using Eq.~(\ref{eqn:RN}) for $N = 1$ (i.e. $r$), $N=3$, and $N=9$, is compared to that for a semi-infinite array, $R_{\infty}$, calculated using Eq.~(\ref{eqn:R}). We assume $\alpha = 10^{-2}$ and $\mu_{0}H_{\mathrm{B}} = 50$mT. The solid vertical line indicates $\omega_{\mathrm{FMR}}$, and the inset is a zoom into the region of the magnetoacoustic resonance, marked by a gray box in the main panel.}
\end{figure} 
%[Why exploring metamat and why around BGS] Magnetoelastic
The metamaterial design studied here and shown in Fig.\ref{fig:fig1}(a) is based on the insights obtained from our analysis of the acoustic scattering by isolated thin magnetic slabs.\cite{Latcham_2019} The magnetoelastic coupling within a slab manifests itself as a peak in the frequency dependence of its reflectivity. This peak corresponds to the Kittel frequency of the slab and is therefore controlled by the bias magnetic field. The strength of the coupling between the localised magnetic and propagating acoustic modes is enhanced for an oblique incidence [see Fig.\ref{fig:fig1}(b)]. As a result, for realistic values of the magnetoelastic coupling, $B$, a noticeable effect is achieved for small values of the Gilbert damping, e.g. $\alpha\simeq10^{-3}$.
For practical uses though, this response needs to be enhanced further. One potential route is to slow down the acoustic modes, increasing their interaction time with the magnetic slabs. This could occur in the vicinity of phononic band gaps. The structure factor of a phononic crystal with embedded magnetic slabs 
%(analyzed in the Supplementary Material and 
[see Fig.\ref{fig:fig1}(c)] exhibits a magnetically tunable anticrossing with the usual phononic dispersion. We expect an enhancement of the response of such a structure when the magnetoacoustic resonance leading to this anticrossing is tuned to the proximity of the band gap. 
%One can also think about enhancement of scattering due to the constructive interference that occur in metamaterial, giving rise to the band gaps. This motivates the study of periodic arrays formed by magnetic slabs. Although the magnetoelastic interaction is typically weak, it has been shown to influence transmission and reflection of acoustic waves\cite{Latcham_2019} in finite sized magnetostrictive slabs, subject to the magnetic damping.
%As this analysis was motivated on keeping the size of the active magnetic element small to reduce the overall loss, a balance is struck on this size, between the interaction strength and said loss. 
%Alternative avenues have been explored to mitigate the influence of the damping, such as enhancing the coupling by introducing a small angle to the incident acoustic wave. Arrays of these resonators have not been addressed and may be exploited by the interaction of the tunable, localised magnetoelastic resonance with resonances in the 1D phononic crystal, forming a hybrid magneto-acoustic metamaterial. Of particular interest are the regimes around the phononic band gaps, as the group velocity of the phonons is reduced, this potentially suppresses the size limitations imposed on the interaction strength. 

The 1D array analyzed here is a simple implementation of this design idea.  Its elementary building blocks are thin ferromagnetic slabs of thickness $\delta$, infinite in the $Y-Z$ plane, and separated by nonmagnetic spacer layers of thickness $\delta_{s}$ $(\delta_{s} \gg \delta)$, as shown in Fig.\ref{fig:fig1}(a). The slabs are magnetized by a bias magnetic field $\mathbf{H}_\mathrm{B} = H_{\mathrm{B}}\hat{\bf z}$ and have saturation magnetization $M_{\mathrm{s}}$. The array either contains a finite number, $N$, of magnetic slabs, or is semi-infinite. Let the $n^{\mathrm{th}}$ resonator be situated at $x_{n} = nL$, where $L = \delta+\delta_s$ is the period of the array. As in Ref.~\citen{Latcham_2019}, we assume that the elastic properties of the two materials may differ. Transverse acoustic waves are obliquely incident on the array from the left. The shear stress produced by the waves perturbs the slabs' magnetization, as described by the standard magnetoelasticity theory.\cite{Comstock_1963, Callen_1965, Akhiezer_1968}
%This array is of finite size, containing a total number of $N$ resonators, such that the rightmost points are $x=0$ and $x=NL$ respectively. 
%This is a simple geometry that admits a full analytic solution, and the 1D array and building block are presented in Fig.\ref{fig:fig1}.(a). 

The magneto-acoustic response of finite arrays are characterized by the reflection, $R_{N}$, transmission, $T_{N}$, and absorption, $A_{N}$, coefficients. Using the transfer matrix method,\cite{Markos_2008} these coefficients can be expressed via the reflection, $r$ and transmission, $t$, coefficients in the forward direction and the reflection, $\tilde{r}$, and transmission, $\tilde{t}$, coefficients in the backward direction. The  coefficients~$t$, ${\tilde t}$, $r$, and ${\tilde r}$ exhibit a strong frequency dependence, which is not specified explicitly here but can be found in Ref.\citen{Latcham_2019}. In particular, this dependence features a Fano resonance near the Kittel frequency, as shown in Fig.\ref{fig:fig1}(b).
%In these single resonators amplitude reflection, transmission and absorption from a slab is given by $r, t$ and $a$ respectively as in Ref.\cite{Latcham_2019}.

The oblique incidence geometry adopted here, exhibits a transverse acoustic displacement $\bm{U} = U(x, y, t)\hat{\bf z}$  inside the $n^{\mathrm{th}}$ layer, $(n - 1)L < x < nL - \delta$, of the nonmagnetic material is given by
\begin{equation}
\label{eqn:Uxcases}
    U(x, y, t)= e^{-i \omega t+ik_{y}y}
    \left[
      A_{n}e^{i\phi_{x}} 
    + B_{n}e^{-i\phi_{x}} \right]
    \ , 
\end{equation}
where $\phi_{x}=k_{x}\left[x-(n - 1)L\right]$. 
%Here the finite $R_{N}$ and $T_{N}$ are defined by right and left traveling acoustic waves out of the array, therefore amplitudes $B_{0}$ and $A_{N}$ respectively.
In what follows, we retain only the $x$-dependence of the wave function.
%$e^{-i\omega t}e^{\pm i\phi_{x, y}} \rightarrow e^{\pm i\phi_{x}}$. % and $\theta^{\prime}_{x} = k\left(x-nL-\delta-\delta_{s}\right)$
Hence, $A_{n}$ and $B_{n}$ represent the amplitudes of acoustic modes traveling to the right and to the left in the $n^{\mathrm{th}}$ nonmagnetic layer, respectively. Then, for a wave of unit amplitude incident from the left onto a finite array, we have $A_{0} = 1$, $B_{0} = R_{N}$, $A_{N} = T_{N}$, $B_{N} = 0$.
%Therefore, $A_{n}$ and $A_{n+1}$ represent the amplitudes of acoustic modes traveling to the right in the $n^{th}$ and $(n+1)^{st}$ non-magnetic matrix elements respectively. Modes in the inverse direction are defined by $B_{n}$ and $B_{n+1}$. 
To form the transfer matrix, $M$, for a single period of the array, amplitudes at $x = nL$ and $x = (n+1)L$ can be related via forward ($t$, $r$) and backward ($\tilde{t}$, $\tilde{r}$) transmission and reflection coefficients. 
Indeed, matching incoming and outgoing waves at the $n^{\mathrm{th}}$ slab, we write
\begin{equation}
    \label{eqn:tandreqns}
    \begin{split}
    A_{n+1}\mathrm{exp}(-i\chi_{\theta}) &= tA_{n}+\tilde{r}B_{n+1}\mathrm{exp}(i\chi_{\theta})\ , \\
    B_{n} &= \tilde{t}B_{n+1}\mathrm{exp}(i\chi_{\theta}) + rA_{n}\ ,
    \end{split}
\end{equation}
where $\chi_{\theta} = \omega\delta_{\mathrm{s}}\sqrt{\rho/C}\mathrm{cos}\theta$ is the acoustic phase delay between two neighboring slabs. 
%For a wave of unit amplitude incident from the left %of a finite $N$ array, $A_{0} = 1$, $B_{0} = R_{N}$, %$A_{N} = T_{N}$, $B_{N} = 0$.
%Here we assume a symmetric system in which a single cell consists of a magnetic slab of thickness $\delta$, with equal non-magnetic matrix width $\delta_{\mathrm{s}}$ to the left and right.
The transfer matrix $M$ is then constructed by inverting Eq.~(\ref{eqn:tandreqns}) as
\begin{equation}
\label{eqn:tmatrix}
M=\begin{Bmatrix}
  \left[t - \tilde{r}r\tilde{t}^{-1}\right]\mathrm{exp}\left(i\chi_{\theta}\right)& \tilde{r}\tilde{t}^{-1}\mathrm{exp}\left(i\chi_{\theta}\right)\\
  -r\tilde{t}^{-1}\mathrm{exp}\left(-i\chi_{\theta}\right) & \tilde{t}^{-1}\mathrm{exp}\left(-i\chi_{\theta}\right)\
 \end{Bmatrix}.
\end{equation}
The action of $M$ can be represented by its eigenvalues $\mu_{\pm}$ and eigenvectors. The eigenvalues that solve the characteristic equation $\mu^{2}-2\epsilon\mu+d=0$ are given by $\mu_{\pm} = \epsilon \mp \sqrt{\epsilon^{2} - d}$, where $d \equiv \mathrm{det}M = \mu_{+}\mu_{-}$ and $2\epsilon \equiv \mathrm{Tr}M = \mu_{+}+\mu_{-}$. 
From Eq.~(\ref{eqn:tmatrix}), we find that $d = t/\tilde{t}$ which has absolute value of one.
As usual, we find that the two eigenvalues of~$M$ either both lie on the unit circle $|\mu| = 1$, or
one is inside and the other is outside. In our system, the energy is dissipated due to the Gilbert damping. 
Hence, we can define $\mu_{\pm}$ so that $|\mu_{+}| < 1$, representing the wave propagating to the right.
%[this part likely needs re-wording]
%Solving the characteristic equation allows us to find eigenvalues by $\mu_{\pm} = \epsilon \mp \sqrt{\epsilon^{2} - d}$.
For a finite array of $N$ resonators, the full transfer matrix $M_{N} = M^{N}$ retains the eigenvector basis with eigenvalues $\mu_{\pm}^{N}$. The initial and final state amplitudes are then projected onto a reciprocal of this basis, multiplied by the eigenvalues, and resolved to obtain for the finite array's reflection coefficient 
\begin{equation}
\label{eqn:RN}
    R_{N} = \frac{R_{\infty}\left(1-\mu_{+}^{2N}\right)}{\left(1-\xi\mu_{+}^{2N}\right)},
\end{equation}
where $R_{\infty}$ is the reflection from a semi-infinite array,
%the reflection from a semi-infinite array and $\xi$ is defined to simplify the expression, both may be expressed as
\begin{equation}
\label{eqn:R}
    R_{\infty} = r\left[ \left(t\tilde{t}-r\tilde{r}\right)\mathrm{exp}\left(i\chi_{\theta}\right) - \tilde{t}\mu_{+}^{-1}\right]^{-1},
\end{equation}
and $\xi$ is defined as,
\begin{equation}
    \xi = \frac{\left(t\tilde{t}-r\tilde{r}\right)\mathrm{exp}\left(i\chi_{\theta}\right) - \tilde{t}\mu_{+}}{\left(t\tilde{t}-r\tilde{r}\right)\mathrm{exp}\left(i\chi_{\theta}\right) - \tilde{t}\mu_{+}^{-1}}.
\end{equation}
The transmission coefficient of the finite array is then given by
\begin{equation}
\label{eqn:TN}
    T_{N} = \frac{\left(1-\xi\right)\mu_{+}^{N}}{1-\xi\mu_{+}^{2N}}.
\end{equation}
The absorbance can be found as $A_{N}^{2} = 1 - |R_{N}|^{2} - |T_{N}|^{2}$. We remind the reader that the parameters $\xi$ and $\mu$ depend on the frequency and the phase delay $\chi_{\theta}$.

To illustrate how $R_{N}$ depends on the number of elements in a finite array, we have performed detailed
calculations for an array of nickel slabs (mass density $\rho = 8900~\mathrm {kgm^{-3}}$, magnetoelastic coupling coefficient $B = 8.8~\mathrm {MJm^{-3}}$, shear modulus $C = 76~\mathrm{GPa}$, gyromagnetic ratio $\gamma = 199~\mathrm{GHz T^{-1}}$, saturation magnetization $M_{\mathrm{s}} = 203~\mathrm{kAm^{-1}}$, $\delta =~30~\mathrm{nm}$\cite{Berk_2019, MatDat_2019}) embedded into a silicon nitride matrix ($\rho_{0}  = 3192~\mathrm{kgm^{-3}}$, $C_{0} = 127~\mathrm{GPa}$, $\delta_s = 500~\mathrm{nm}$\cite{Huszank_2015, MatDat_2019} % - replace this with a more appropriate reference
). Fig.\ref{fig:RnFigure} presents results of the calculations for a generic case, without fine-tuning of the magnetoelastic resonance. For $N > 1$, the absolute value of the reflection coefficient has the unity value in frequency regions corresponding to the acoustic stopbands (phononic band gaps). 
%\setcitestyle{square}?
These are caused by the mismatch of the acoustic impedance $Z = \sqrt{\rho C}$ at slabs' surfaces and occur even in the absence of magnetoelastic coupling ($B=0$).\cite{Martin_2015, Born_1964, Brekhovskikh_1997}
%(for a situation in which the elastic mismatch is removed, a similar band gap spectrum similar is found due to the periodic impedance mismatch caused by the magneto-elastic coupling\cite{}).
%In these stopbands $|\epsilon| > 1$,  while the passbands that form the transmission band occur when $|\epsilon| < 1$ . 
Each passband contains $N-1$ peaks, which are due to the phase delay of the acoustic waves increasing by $\pi$ across each Brillouin zone.\cite{Markos_2008} The magnetoelastic coupling ($B \neq 0$) manifests itself via an asymmetric peak due to the Fano resonance, positioned at the Kittel frequency $\omega_{\mathrm{FMR}} = \gamma\mu_{0}\sqrt{H_{\mathrm{B}}(H_{\mathrm{B}}+M_{\mathrm{s}})} \simeq 6.7~\mathrm{GHz}$ at $\mu_{0}H_{\mathrm{B}} = 120~\mathrm{mT}$.\cite{Kittel_1958}
%The frequency dependence of $T_{N}$ and $A_{N}$ are given in supplementary text: $T_{N}$ mirrors $R_{N}$ aside from near the Fano resonance region, where the linewidth of $A_{N}$ increases with $N$.  

The rapid oscillation in passbands in Fig.\ref{fig:RnFigure} is due to the multiple reflections within an array of finite size. For sufficiently large arrays (i.e. when the decay length is smaller than the array size), these oscillations are suppressed. Indeed, the oscillations are suppressed for $R_{\infty}$ (calculated using Eq.~(\ref{eqn:R}) and shown by the solid line in Fig.\ref{fig:RnFigure}), as expected for $N \rightarrow \infty$. So, our subsequent analysis is focused on the semi-infinite array. 
%In the limit $N \rightarrow \infty$, $|T_{N}| \rightarrow 0$ for all $f$, and energy is either only attenuated $A_{\omega}$, or reflected $R_{\omega}$, such that 1 = $|A_{\omega}|^{2}+|R_{\omega,}|^2$. In this regime Eqn.~(\ref{eqn:RN}) simplifies to Eqn.~(\ref{eqn:R}).
%For a completely lossless ($B=0$) system the scattering from this semi-infinite array would have to be defined by an alternative approach\cite{Martin_2015}, but due to the finite magnetic damping the rapid oscillation seen in the passband is averaged. 
\begin{figure}[t!]
    \centering
    \includegraphics[width=85mm]{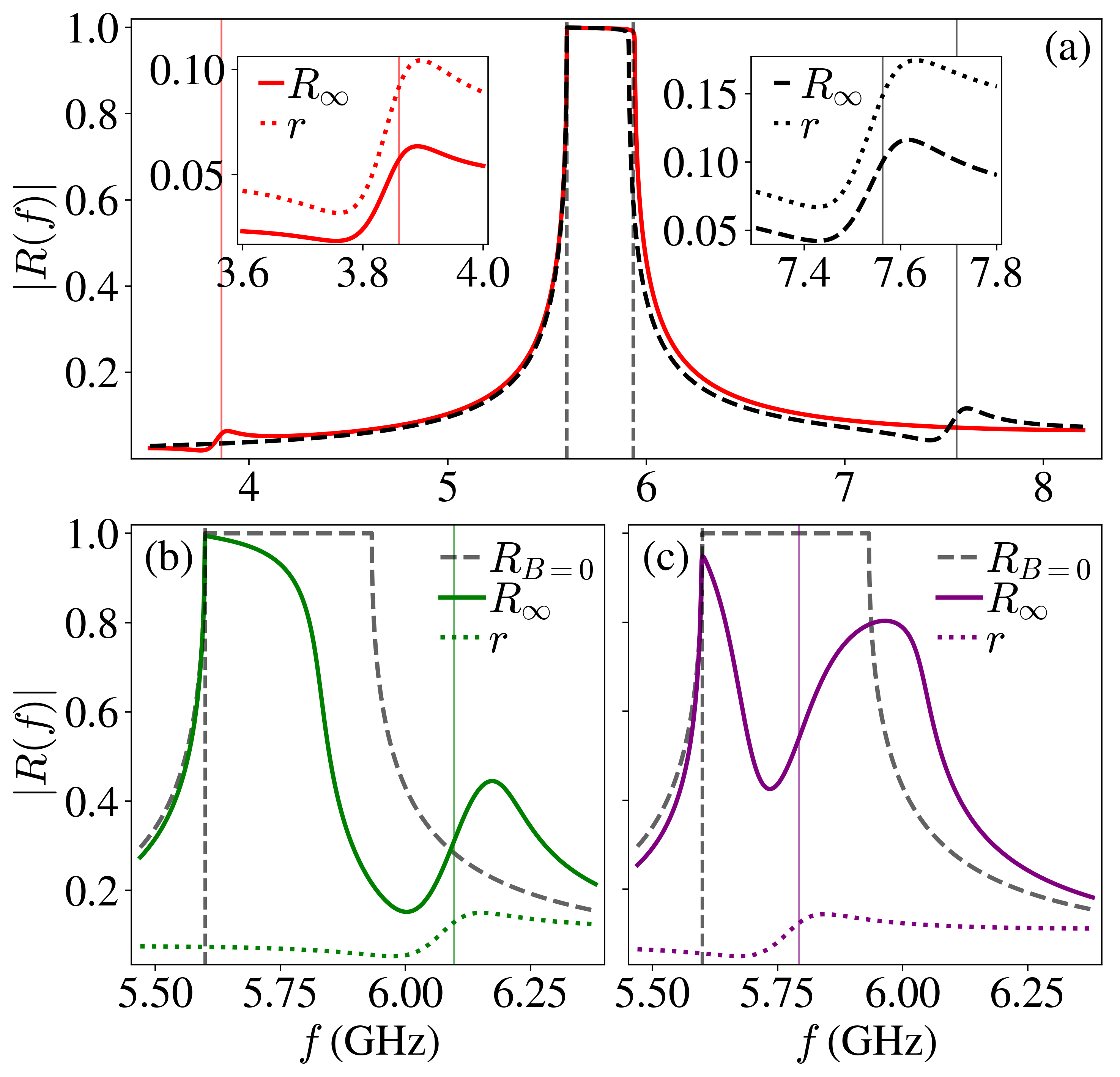}
    \caption{\label{fig:RinfFigure}The three regimes identified for the tuning of the Kittel frequency $\omega_{\mathrm{FMR}}$ by $\mu_{0}H_{B}$ in the semi-infinite array. For all, $\alpha = 10^{-2}$, dashed vertical lines indicate the position of the band gap frequencies and solid vertical lines indicate the position of $\omega_{\mathrm{FMR}}$. A dashed black curve is also shown for $R_{B=0}$ which represents $R_{\infty}$ with $B=0$. (a) Regime I, $\omega_{\mathrm{FMR}}$ is tuned away from the phononic band gap. Insets compare $R_{\infty}$ (solid, Eqn.~(\ref{eqn:R})) with $r$\cite{Latcham_2019} (dotted) at $\mu_{0}H_{\mathrm{B}} = 50$mT (left, solid red) and $145$mT (right, dashed black). (b) Regime II, $\omega_{\mathrm{FMR}}$ is tuned close to the band frequencies using $\mu_{0}H_{\mathrm{B}} = 105$mT. (c) Regime III, $\omega_{\mathrm{FMR}}$ is tuned inside the band gap using $\mu_{0}H_{\mathrm{B}} = 97$mT.}
\end{figure}

Fig.\ref{fig:RinfFigure} displays the reflectivity $R_{\infty}$, of a semi-infinite array over frequency for several positions of the Kittel resonace. 
%As the main feature of an uncoupled phonon scattering spectrum is the opaque nature at band gap frequencies, as seen in the grey curve in all panels of Fig.\ref{fig:RinfFigure}, thus 
We identify three regimes based on the positioning of $\omega_{\mathrm{FMR}}$  relative to phononic band gaps.
%The strength of the magnetoelastic scattering depends on $\omega_{\mathrm{ME}}$, as its position on the background acoustic reflectivity landscape. It is possible to identify three regimes in this regard, as detailed in the three panels of Fig.\ref{fig:RinfFigure}. 
Regime I (`detuned') occurs as $\omega_{\mathrm{FMR}}$ is tuned inside a passband, away from band edges. Shown by Fig.\ref{fig:RinfFigure}.(a),  with insets providing comparison of $R_{\infty}$ with $r$. In both cases (tuned above or below), the peak 
in~$R_{\infty}$ is lower than that in~$r$. 
This suppression is caused by the destructive interface of backward reflected waves, which 
is pronounced away from band edges.
%This discrepancy is caused by the suppression of multiple coherent reflections by the attenuation. In this frequency region the uncoupled ($B=0$) acoustic reflectivity is less than the Fabry-P\'erot resonance of a single resonator.
%This is found as for a finite or infinite array the average background acoustic reflectivity is less than the Fabry-P\'erot resonance associated with a single inclusion.

Regime II (`adjoining') occurs as the Kittel frequency $\omega_{\mathrm{FMR}}$, approaches the band gap from a passband. 
%Here the unccoupled ($B=0$) background reflectivity $R_{B=0}$ is larger than the Fabry-P\'erot resonance of $r$, as seen in Fig.\ref{fig:RinfFigure}.(b). The background $R_{B=0}$ is given here for comparison, as the influence of magnetoelastic modes skew the lower band edge. 
Here the resonant scattering becomes highly sensitive to the detuning of $\omega_{\mathrm{FMR}}$ from the band edge. In close proximity to the band gap, where
the Bragg condition holds, the scattering is enhanced.
(In other words, destructive interference crosses
over to constructive.)
Inside the band gap the reflectivity is reduced, which may be interpreted as induced transparency. 
%To confirm this interpretation for a finite $N$ array, $T_{N}$ is shown in supplementary text.
This reduction in reflectivity is caused by the magnetoacoustic hybridization providing slowly propagating hybrid modes inside the band gap. Indeed, Kittel resonances in the magnetic elements may hybridise via `virtual phonons' in the stop band. This would introduce a non-zero density of states inside the band gap which can also be seen as coherent reflections in a finite array.
%Which we consider by noting the presence of a non-zero density of states as $S(f, k)$ enters the band gap and by observing that the coherent oscillations discussed previously return for a finite $N$ array.
%reflectivity is suppressed. This manifests due to the reduced group velocity $v_{\mathrm{g}}$, at these frequencies the acoustic waves spend more time in the absorbing elements, increasing loss. 

Regime III (`inner') occurs as the Kittel frequency $\omega_{\mathrm{FMR}}$, is tuned inside the band gap. The main feature of this regime is the reduced reflectivity, as seen in Fig.\ref{fig:RinfFigure}.(c). This reduction is not symmetric as the bias field sweeps the Kittel resonance over the band gap.
We see that the behaviour at the upper and the lower
edge is distinctly different: the reflectivity is reduced as $\omega_{\mathrm{FMR}}$ approaches the upper band gap frequency. This phenomenon can be attributed to the Borrmann effect.\cite{Campbell_1951, Batterman_1964} In a pure phononic crystal ($B=0$) the modes at the band edges are two standing waves, phase shifted by $90^{\circ}$.\cite{Croenne_2011} For one of the modes the maxima of the stress occur in magnetic slabs, while for the other this pattern is reversed: the slabs become the nodes. When the modes are coupled to the magnetisation dynamics in the slabs, the dissipated energy depends primarily on the local 
stress. Hence absorption is suppressed for the latter, nodal mode.\cite{Novikov_2017} 
This mode occurs at the lower band edge
if the acoustic impedance of the magnetic (M) material is larger than that of the non-magnetic matrix (NM): $Z_{\mathrm{M}}> Z_{\mathrm{NM}}$. This gives 
rise to the asymmetry seen in Fig.\ref{fig:RinfFigure}.
%Z_{\mathrm{M}}$). 
%The maxima is localised in magnetic slabs at the %lower band edge when the acoustic impedances of the %magnetic (M) material is smaller than the %non-magnetic (NM) ($Z_{\mathrm{NM}}> 
%Z_{\mathrm{M}}$). 
%Absorption is therefore suppressed at the upper band %edge, illustrated as an asymmetric response in %reflectivity.\cite{Novikov_2017} 
(The asymmetry is reversed when $Z_{\mathrm{M}} < Z_{\mathrm{NM}}$.)
%Thus, at the lower band edge frequency For the relation of acoustic impedances in the magnetic (M) and non-magnetic (NM) materials is $Z_{\mathrm{NM}}> Z_{\mathrm{M}}$, as in Fig.\ref{fig:RinfFigure}, 
%Calculations of impedance show the lower is localised in the magnetic and upper non-magnetic materials (as $Z_{\mathrm{NM}} > Z_{\mathrm{M}}$), thus the assymmetry can be seen therefore as a reduction of the magnetoelastic coupling when energy is localised in the non-magnetic matrix. This leads to the reduction of absorption of the propagating hybrid modes. With the particular parameters presented this occurs at the upper band frequency, but this may reversed if acoustic impedances are adjusted ($Z_{\mathrm{M}} > Z_{\mathrm{NM}}$).

We note that magneto-elastic effects shown in Fig.\ref{fig:RinfFigure} remain significant 
%size for hybrid magneto-acoustic metamaterials, 
even for a realistic damping value $\alpha=10^{-2}$. This is a considerable improvement compared with a single resonator where a similar damping value would completely suppress the Fano features.\cite{Latcham_2019} 
A shift in lower band edge [see Fig.\ref{fig:RinfFigure}.(a)] is induced by proximity to the Kittel frequency $\omega_{\mathrm{FMR}}$. This band shift and the induced transmission are resolved separately as $\omega_{\mathrm{FMR}}$ sweeps the band gap when the band gap width significantly exceeds the Fano resonance linewidth.

\begin{figure}[t!]
    \centering
    \includegraphics[width=85mm]{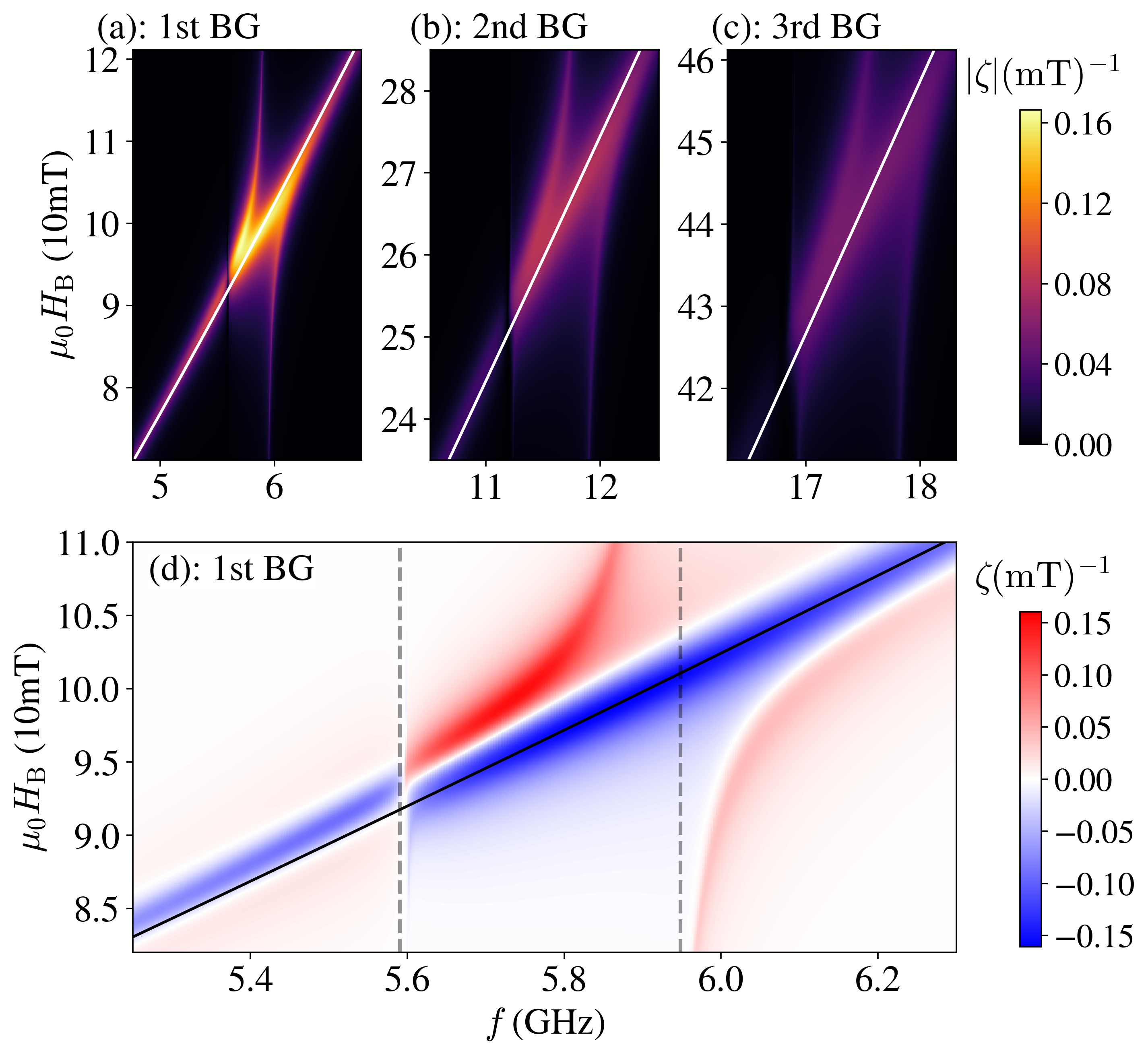}
    \caption{\label{fig:ModulationFig} The modulation coefficient $\zeta = \partial |R_{\infty}|/\partial H_{\mathrm{B}}|$, is illustrated over the first three (a) first, (b) second, (c) third phononic band gaps, by varying with $\mu_{0}H_{\mathrm{B}}$ and $f$ with $\alpha = 10^{-2}$. The linear, solid while line represents $f_{\mathrm{ME}}$ in each panel. (d) A focus around the first band gap, retaining the phase inversion while the Kittel resonance traverses the band gap. The position of the band gap edges at $B=0$ are marked with dashed vertical lines.}
\end{figure}

To characterise the tunability of the magnetoelastic resonance we introduce the modulation coefficient $\zeta = \partial |R_{\infty}|/\partial H_{\mathrm{B}}$ showing the variation of $R_{\infty}$ by $\mu_{0}H_{\mathrm{B}}$. As $\omega_{\mathrm{FMR}}$ is tuned around band gaps the reflectivity becomes sensitive to changes in the external bias field $\mu_{0}H_{\mathrm{B}}$, as illustrated by $\zeta$ in Fig.\ref{fig:ModulationFig}.(a). Here, the first three band gaps show this phenomena occuring at higher band frequencies. However, operating a device 
%around these regions 
in this high-frequency regime may be impractical as e.g. it would require a large bias field ($>0.5$T).
%Higher frequencies are not illustrated as the exchange field no longer becomes negligable and $\mu_{0}H_{\mathrm{B}}$ becomes too large (>1T). Indeed, if one considers the third band edge, the associated exchange and magnetodipolar fields are $\mu_{0}M_{s}(kl_{ex})^{2} \simeq 218$mT and $\mu_{0}M_{s}k_{y}\delta \simeq 115$mT respectively, which are still small compared with the large bias $\mu_{0}H_{\mathrm{B}} \simeq 620$mT.
%The modulation of $\zeta$ predominantly occurs around band gaps, so this is where we focus our attention. By neglecting 
%Here, successive band gaps increase in width. We expect this to improve the resolution between features at higher magnetic damping $\alpha$. 
%As such 
Fig.\ref{fig:ModulationFig}.(d) provides an insight into the behaviour in the first band gap. 
%illustrates the first band gap. 
%Here $R_{\infty}$  
We see that~$\zeta$ is enhanced in a close proximity
to the band gap edges (dashed gray) and the Kittel frequency (solid black), rapidly changing from 
positive (red) to negative (blue). 
%is shown to increase (red) and decrease (blue) in %separate regions around the Kittel frequency (black, %solid) and uncoupled band edges (grey, dashed). 
Frequencies around the Kittel resonance are either increased or decreased in frequency by coupling to spin wave modes [see Fig.\ref{fig:fig1}.(c)]. Band edges are influenced by this shift in frequency, relative to their own position with respect to the magneto-elastic anti-crossing. For larger band widths and closer proximity to the Kittel frequency this shift is enhanced. At the lower band edge, the Borrmann effect counteracts the shift [see Fig.\ref{fig:RinfFigure}.(a)].
The Borrmann effect also results in the asymmetry below the lower band edge ($\simeq 5.6$GHz), as the Fano resonance reconstructs its lineshape around $\omega_{\mathrm{FMR}}$ when exiting the bad gap. 

%Fig.\ref{fig:ModulationFig}.(b) shows the quick modulation of reflectance and absorbance in frequency. As suppression of reflectivity is observed alongside gain inside the phononic crystal, for large $N$, $T_{N} \rightarrow 0$ therefore more energy is lost to the lattice. This suggests any acoustic wavepacket tuned inside the gap could be decomposed in frequencies, with trailing frequencies at any $H_{\mathrm{B}}$ being predominantly reflected rather than absorbed, opening up potential uses in acoustic devices.

In summary we have shown that the metamaterial approach is indeed helpful for magnetoacoustics.
%more is indeed different.\cite{Anderson_1972} 
Hybrid metamaterials, formed by 1D arrays of resonators, magnify the effects of magnetoelastic coupling, thus mitigating the Gilbert damping to tolerable levels. The structures considered here are tunable by an applied bias field and exhibit a rich and complex behaviour, such as induced transmission and Borrmann asymmetry. To prototype realistic structures, we aim to investigate higher dimensions and implement surface acoustic waves. We envision the characteristics shown here will prove useful when engineering sensors, actuators, radio frequency modulators and reconfigurable magnonic devices.
%In summary we have demonstrated that arrays of 1D resonators form tunable 1D metamaterials. We show that the reflection and transmission from such a structure can be controlled by the external bias field...
%Note increase in resonant scattering - sensitivity to bias field around band edges
%induced transmission caused by coupling
%Borrmann effect causes asymmetry in band edges
%Something about non-uniformity increasing asymmetry

%Supplementary material is available at: ...
%See the supplementary material for derivations and analysis of the spectral function $S(f, k)$, derivation of the reflectivity from a semi-infinite array using an elementary method in the case of a lossless system ($B=0$) and ($T_{N}, A_{N}$) graphs.

The research leading to these results has received funding from the Engineering and Physical Sciences Research Council of the United Kingdom (Grant No. EP/L015331/1) and from the European Union’s Horizon 2020 research and innovation program under Marie Skłodowska-Curie Grant Agreement No. 644348 (MagIC).  

\bibliography{MagnonPhononMetamat}

\end{document}